\documentstyle[prd,preprint,aps,epsfig]{revtex}
\begin{document}

\draft
\def\be{\begin{equation}}
\def\ee{\end{equation}}
\def\bea{\begin{eqnarray}}
\def\eea{\end{eqnarray}}
\def\nn{\nonumber}
\def\ep{\epsilon}
\def\c{\cite}
\def\m{\mu}
\def\ga{\gamma}
\def\lan{\langle}
\def\ran{\rangle}
\def\Ga{\Gamma}
\def\thet{\theta}
\def\la{\lambda}
\def\Lam{\Lambda}
\def\ka{\chi}
\def\si{\sigma}
\def\al{\alpha}
\def\pa{\partial}
\def\de{\delta}
\def\De{\Delta}
\def\Dex{\Delta_{,x}}
\def\Dey{\Delta_{,y}}
\def\Dev{\Delta^{-1}}
\def\Ome{\Omega}
\def\Om2{\Omega^{2}}
\def\ov{\over}
\def\gmn{g_{\mu\nu}}
\def\gmnv{g^{\mu\nu}}

\def\rsr{{r_{s}\over r}}
\def\rrs{{r\over r_{s}}}   
\def\rs2r{{r_{s}\over 2r}}
\def\l2r2{{l^{2}\over r^{2}}}
\def\rsa{{r_{s}\over a}}
\def\rsb{{r_{s}\over b}}
\def\rsro{{r_{s}\over r_{o}}}
\def\rss{r_{s}}
\def\a2{{l^{2}\over a^{2}}}
\def\b2{{l^{2}\over b^{2}}}
\def\op{\oplus}
\def\sn{\stackrel{\circ}{n}}
\def\c{\cite}

%\twocolumn[\hsize\textwidth\columnwidth
%\hsize\csname @twocolumnfalse\endcsname

\title
{Axial Torsion-Dirac spin  Effect in Rotating Frame with Relativistic Factor}
\author{C. M. Zhang } 
%\author{xxxxxxxxx Draft xxxxxxx } 
\address{
   Research Center for
       Theoretical Astrophysics\\
       School  of  Physics,
       University of Sydney,
       NSW 2006, Australia\\
    zhangcm@physics.usyd.edu.au}

%\date{\today}

\maketitle

\begin{abstract}

In the framework of spacetime with torsion and without curvature, 
the Dirac particle spin precession in the rotational system is studied. 
 We write out  the  equivalent tetrad of rotating frame, in the polar 
coordinate system,  through  considering the relativistic 
factor,  and the resultant   equivalent  metric is a flat  Minkowski one. 
The obtained rotation-spin coupling formula can be
applied to the 
high speed rotating case, which  is consistent with the expectation.

\end{abstract}


{\bf key words}: torsion, Dirac particle, rotation-spin(1/2), noninertial effect

%\vskip1pc

\vskip1pc]

%\newpage

\section{Introduction}

The tetrad theory of gravitation has been pursued by a number of authors 
~\cite{hay79,nh80,per1,per2,perbook,pvz01,mal},  
 where the spacetime  is characterized by the
 torsion tensor and the vanishing  curvature, the relevant spacetime is the
Weitzenb\"ock spacetime~\cite{hay79}, which is a special case of the Riemann-Cartan
spacetime with the constructed  metric-affine theory of
 gravitation \c{h76,h91,hmpd}. 
 The tetrad theory of gravitation  will be equivalent to general relativity
 when the convenient choice of the parameters of  the Lagrangian.

We will use the greek alphabet ($\mu$, $\nu$, $\rho$,~$\cdots=1,2,3,4$)
to denote tensor indices, that is, indices related to spacetime. The latin alphabet
($a$, $b$, $c$,~$\cdots=1,2,3,4$) will be used to denote local Lorentz (or tangent space)
indices. Of course, being of the same kind, tensor and local Lorentz indices can be
changed into each other with the use of the tetrad $e^{a} {}_{\mu}$, which satisfy
\be
e^{a}{}_{\mu} \; e_{a}{}^{\nu} = \delta_{\mu}{}^{\nu} \quad
; \quad e^{a}{}_{\mu} \; e_{b}{}^{\mu} = \delta^{a}{}_{b} \; .
\label{orto}
\ee
A nontrivial tetrad field can be used to define the linear Cartan connection\c{hay79,perbook}
 
\be
\Gamma^{\sigma}{}_{\mu \nu} = e_a{}^\sigma \partial_\nu e^a{}_\mu \;,
\label{car}
\ee 
with respect to which the tetrad is parallel:  
\be {\nabla}_\nu \; e^{a}{}_{\mu}
\equiv \partial_\nu e^{a}{}_{\mu} - \Gamma^{\rho}{}_{\mu \nu}
\, e^{a}{}_{\rho} = 0 \; . 
\label{weitz}
\ee 
The Cartan connection can be decomposed according to 
\be
{\Gamma}^{\sigma}{}_{\mu \nu} = {\stackrel{\circ}{\Gamma}}{}^{\sigma}{}_{\mu
\nu} + {K}^{\sigma}{}_{\mu \nu} \; ,
\label{rel} 
\ee
where
\be
{\stackrel{\circ}{\Gamma}}{}^{\sigma}{}_{\mu \nu} = \frac{1}{2}
g^{\sigma \rho} \left[ \partial_{\mu} g_{\rho \nu} + \partial_{\nu}
g_{\rho \mu} - \partial_{\rho} g_{\mu \nu} \right]
\label{lci}
\ee
is the Levi--Civita connection of the metric

\be
g_{\mu \nu} = \eta_{a b} \; e^a{}_\mu \; e^b{}_\nu \; ,
\label{gmn}
\ee
 where $\eta^{ab}$ is the metric in flat space with the line element

\be
d\tau^{2} = g_{\mu \nu} dx^{\mu} dx^{\nu} \;,
\ee 

and 

\be {K}^{\sigma}{}_{\mu \nu} = \frac{1}{2}
\left[ T_{\mu}{}^{\sigma}{}_{\nu} + T_{\nu}{}^{\sigma}{}_{\mu}
- T^{\sigma}{}_{\mu \nu} \right]
\label{conto}
\ee 
is the contorsion tensor, with 
\be
T^\sigma{}_{\mu \nu} =
\Gamma^ {\sigma}{}_{\mu \nu} - \Gamma^{\sigma}{}_{\nu \mu} \;  \label{tor} 
\ee
the torsion of the Cartan connection~\c{hay79,perbook}. The irreducible
torsion vectors, i.e., the torsion vector and 
the torsion axial-vector, can then be
constructed as~\c{hay79,perbook} 
\be
V_{\mu} =  T^{\nu}{}_{\nu \mu}\,,
\ee
\be
A_{\m} = {1\over 6}\ep_{\m\nu\rho\si}T^{\nu\rho\si}\,,
\ee
with $\ep_{\mu \nu \rho \sigma}$ being the completely antisymmetric
tensor normalized as $\ep_{0123}=\sqrt{-g}$  
and $\ep^{0123}=\frac{1}{\sqrt{-g}} $.

The spacetime dynamic effects on the spin is  incorporated into Dirac
equation  through the ``spin connection''  appearing in the Dirac equation
in  gravitation \cite{hay79}. 
 In Weitzenb\"ock spacetime, as well as the general version of torsion 
gravity, it has been shown by many
authors~\c{hay79,nh80,ham94,ham95,heh71,tra72,rum79,yas80,aud81} that the spin
precession of a Dirac particle  is intimately
related to the torsion axial-vector,
 and it is interesting to note  that the 
torsion  axial-vector represents the deviation of the axial symmetry
from the spherical symmetry \cite{nh80}.

\be
\frac{d{\bf S}}{dt} = - {3\ov2} \mbox{{\boldmath $A$}} \times {\bf S}, 
\label{precession1}
\ee
where ${\bf S}$ is the semiclassical spin vector of a Dirac particle, 
and $\mbox{{\boldmath $A$}}$ is the spacelike part 
of the torsion axial-vector.
Therefore, the corresponding extra Hamiltonian energy is of the form, 

\be
\de H = - {3\ov2} \mbox{{\boldmath $A$}}\cdot \mbox{\boldmath$S$}\,.
\label{ham2}
\ee

Throughout this paper we use the relativistic unit, $c=1$.

\section{The rotation-spin effect}

Now we discuss the  Dirac equation in the 
rotational coordinate system  with the polar  coordinates 
$(t,r,\phi,z)$, and the system is rotating  with the 
angular velocity $\Ome$,  and  the rotation axis is set in z-direction.

In the case of considering the relativistic factor $\ga = 1/\sqrt{1-(\Ome r)^{2}}$, 
the tetrad can be expressed by the 
dual basis of the differential one-form \c{hn} through  
choosing a coframe of the rotational coordinate  system,  so we define, 

\bea
  d\vartheta ^{\hat{0}}& =& \,  \ga[d\,t - \Ome r (rd\phi)]\;, \\
\label{coframe0}
\quad d\vartheta ^{\hat{1}} &=&\, dr\;,\\
\quad d\vartheta ^{\hat{2}} &=&\,  \ga[ (rd\phi) - \Ome r dt]\;,\\
\quad d\vartheta ^{\hat{3}}&=& \, dz \;,
\label{coframe}
\eea
however our result here is added by the relativistic factor $\ga $ because the high 
speed rotation is taken into account. If $\Ome r $ is much less than the speed of light, 
 then we have  $\ga = 1 $  and the classical coframe expression is recovered, 
which is same as those applied in the existed references \c{hn,zcm2002}. 
Therefore Eq.(\ref{coframe0}) and Eq.(\ref{coframe}) 
is a generalised coframe expression for any rotation  velocity.

The tetrad can be obtained with the subscript 
$\mu$ denoting the column index (c.f. \c{hn,zcm2002}),  

\be
e^{a}{}_{\mu} = \pmatrix{
\ga    & 0 &  -\ga \Ome r^{2}& 0 \cr
0 & 1 & 0 & 0 \cr
-\ga \Ome r & 0 & \ga  r & 0 \cr
0      & 0 & 0 & 1} ,
\label{te1}
\ee 
with the inverse 
$e_{a}{}^{\mu} = \gmnv e^{b}{}_{\nu} \eta_{ab}$, and so
\be
e_{a}{}^{\mu} = \pmatrix{
\ga   & 0  & \ga \Ome  & 0 \cr
0 & 1 &0  &0\cr
\ga \Ome r &0 &\ga/r  &0 \cr
0 &0 &0  &1} .
\label{te2}
\ee

We can inspect that 
Eqs.(\ref{te1}) and (\ref{te2}) satisfy the conditions in 
Eqs.(\ref{orto}) and (\ref{gmn}).     

The obtained metric is obtained  as 

\be
\gmn {} = \pmatrix{
1  & 0 &  0 & 0 \cr
0  & -1& 0       & 0 \cr
0  & 0      & -r^{2}       & 0 \cr
0      &0       & 0       &-1} ,
\ee 

\be
\gmnv {} = \pmatrix{
1  & 0 &  0 & 0 \cr
0  & -1& 0       & 0 \cr
0  & 0      & -1/r^{2}       & 0 \cr
0      &0       & 0       &-1} ,
\ee 
and with the determinant of the metric
\be
g=det|\gmn|= -r^{2}\;.
\ee
From the tetrad and metric given above, we have the line element, 
\bea
d\tau^2 &=& \eta_{ab}\,d\vartheta^{a} \otimes d\vartheta^{b} = g_{\mu\nu}dx^{\mu}dx^{\nu} \nn \\
      &=&  dt^{2} - (dr^{2} + r^{2}d\phi^{2} + dz^{2}  )\;, 
\label{dsr}
\eea
 and we find that the metric in Eq.(\ref{dsr}) is that of flat spacetime, which will result in 
the null curvature. Although the  curvature vanishes, the torsion (field) may have  nonzero
 components determined by tetrads and  not by metrics.
 In other words, the basic element in torsion gravity without curvature is
tetrad and the metric is just a by-product \c{hay79}.  
{}From Eqs.(\ref{te1}) and (\ref{te2}), we can now construct the
Cartan connection, whose nonvanishing components are:
\be
\Ga^{0}{}_{01} = \ga^{2}  \Ome^{2} r,  \;\; \;\;
\Ga^{2}{}_{01} = - \ga^{2}  \Ome/r, \;\;
\Ga^{0}{}_{21} =- \ga^{2}  \Ome r, \;\;
\Ga^{2}{}_{21} = \ga^{2}/r\;.
\ee

 The corresponding nonvanishing torsion components 
contributed to the   axial torsion-vectors  are:

\be
T^{2}{}_{01} = - \ga^{2}  \Ome/r,   \;\;
T^{0}{}_{21} =  - \ga^{2}  \Ome r\;\;, 
\ee

The  nonvanishing axial torsion-vectors  are consequently

%\be
%V_{\m}  = 0,   \;\; \m=0,1,2,3\;, 
%\ee

\be
A_{3} = {2\ov 3}\ga^{2} \Ome \; ,   \;\; A_{k} = 0, k=0,1,2\;.
\ee

As shown, $ A_{1}  = A_{2} = 0$ is  on account of the Z-axis
symmetry which results in the cancelling of the $r$ and $\phi$  
components, and then generally we can write 
$\mbox{{\boldmath $A$}} = {2\ov3}\ga^{2}{\bf \Ome}$.
 From the spacetime geometry view, 
the torsion axial-vector represents the deviation from the spherical
symmetry~\c{nh80}, i.e., which will  disappear in the spherical case 
(Schwarzschild spacetime for instance) and occurs in the axisymmetry 
case (Kerr spacetime for instance). 
Therefore the torsion axial-vector 
corresponds to an inertia field with respect to Dirac particle, 
which is now explictly expressed by Eq.(\ref{precession1}), 

\be
\frac{d{\bf S}}{dt} = - \ga^{2} \mbox{{\boldmath $\Ome$}} \times {\bf S}\;.
\label{precession2}
\ee
If the physics measurement is performed in the rotating frame,  the time dt 
is taken as the proper time through setting the null space difference. Then   
we have  $dt = d\theta^{0}/\ga$, and so 

\be
\frac{d{\bf S}}{d\theta^{0}} = - \ga \mbox{{\boldmath $\Ome$}} \times {\bf S}\;,
\label{precession3}
\ee
 and the additive Hamiltanian energy measured in the rest frame is,  

\be
\de H = - \ga \mbox{{\boldmath $\Ome$}}\cdot \mbox{\boldmath$S$}\;, 
\label{ham3}
\ee
 which is same as that expected by Mashhoon (c.f. Ref.\c{mas88,mas20}).

\section{Discussions and conclusions}
The rotation  Dirac spin coupling for the high speed rotation in the framework of the 
 torsion  spacetime without curvature has been derived. 
This effect was first proposed by Mashhoon\c{mas88,mas20}, and 
the straightforward theoretical derivation was performed by Hehl and Ni 
\c{hn}. However, the relativistic factor has not been considered in the 
 previous theoretical work on rotation-spin \c{hn,zcm2002}. So, in this paper, 
we follow the axial torsion spin treatment \c{zcm2002}, and extend  that method 
to the high speed case, which successfully presents the relativistic $\ga$ 
factor  into the rotation-spin coupling term.  One fact seems to be interested 
to be paid attention that the choice of the tetrad results in the flat metric, 
which produces the null curvature, i.e., Riemannian curvature and Cartan curvature. 
It is remarked that the flat metric will 
arise the Minkowski spacetime, not Riemannian spacetime, but the axial torsion 
has nothing to do with the metric and is just related to the tetrad. Technically, 
the diagonal tetrad will produce the diagonal metric, but the inverse is not true. 
So our  non-diagonal tetrad, similar to  Lorentz transformation, arised from the 
rotation velocity and  relativistic fator,  results in
the diagonal metric. In other words, the metric determines the curvature of spacetime, 
 which reflects the gravitation and define the geodesic of free particle, then 
tetrad or torsion, determining the metric as well, defines the rotation-spin 
motion. Although our final conclusion is same as that by Hehl and Ni\c{hn}, 
they obtained 
the non-diagonal metric, which was derived from the Fermi-Walker transport, i.e., 
non-Minkowski spacetime. The relation between the geometrical meaning and physical 
meaning  has not yet been clearly and needs the further investigation.     
 Nontheless, the axial torsion-spin coupling is also successfully applied to 
the Kerr spacetime\c{pvz01},  and  the gravitomagnetic effect on Dirac particle 
\c{mas20} has been obtained, where the exact Kerr tetrad has been exploited. 
This means that  the axial torsion-spin method is applicable to the axisymmetric 
 interaction effect on the Dirac particle.

\section*{Acknowledgments}
Discussions and suggestions, as well as critic reading,  from G. Lambiase    
 are highly appreciated.

\end{document}